\documentclass[twocolumn,abs,prb,showpacs]{revtex4-1}

\usepackage{graphicx}
\usepackage{dcolumn}
\usepackage{float}
\usepackage{amsmath}
\usepackage{bm}

\newcommand{\comment}[1]{}

\begin{document}


\title{AC/DC Spin and Valley Hall Effects in Silicene and Germanene}

\author{C.J. Tabert}
\author{E.J. Nicol}
\affiliation{Department of Physics, University of Guelph,
Guelph, Ontario N1G 2W1 Canada} 
\affiliation{Guelph-Waterloo Physics Institute, University of Guelph, Guelph, Ontario N1G 2W1 Canada}
\date{\today}

\begin{abstract}
{The intrinsic spin and valley Hall conductivities of silicene, germanene and other similar two dimensional crystals are explored theoretically.  Particular attention is given to the effects of the intrinsic spin-orbit coupling, electron doping and the type of insulating phase of the system (\emph{i.e.}, a topological insulator or a band insulator) which can be tuned by a perpendicular electric field.  At finite frequency, the transverse edge to which carriers of particular spin and valley label flow can be controlled such that an accumulation of a particular combination of spin and valley index can be obtained.  The direction of flow is found to be dependent on the type of insulating phase.  The magnitude of the Hall conductivity response is enhanced from the DC values at certain incident photon frequencies associated with the onset of interband transitions.  Analytic results are presented for both the DC and finite frequency results.
}
\end{abstract}


\maketitle

\section{Introduction}

Silicene, the monolayer of silicon isostructural to graphene, has gathered increasing attention as it has recently been synthesized\cite{Lalmi:2010, DePadova:2010, DePadova:2011, Vogt:2012, Lin:2012, Fleurence:2012} and, unlike graphene, has a sizable spin orbit interaction making it more susceptible to spin manipulation.  The spin orbit interaction is predicted to open a gap of order  $\Delta_{\rm so}\approx 1.55-7.9$meV\cite{Liu:2011, Liu:2011a, Drummond:2012} in the low-energy Dirac-like band structure and, as a result, it has been argued that silicene is a topological insulator (TI).  Due to the buckled structure of the lattice, it also displays a tunable band gap\cite{Drummond:2012} in the presence of a perpendicular electric field which induces a transition from a TI to a band insulator (BI)\cite{Drummond:2012,Ezawa:2012a}.  Furthermore, unlike graphene, recent density-functional band-structure calculations have predicted that silicene should exhibit a quantum spin Hall effect at an accessible temperature\cite{Liu:2011}.   These properties, along with silicene's compatibility with the existing silicon-based nanoelectronic industry, make silicene a promising material for technological applications.  Germanene, a monolayer of germanium, has yet to be fabricated; however, it is analogous to silicene but with a much larger predicted spin-orbit gap of $\Delta_{\rm so}\approx 24-93$meV\cite{Liu:2011, Liu:2011a}. The results discussed below apply equally to this system and to other two-dimensional (2D) systems with the same low-energy Hamiltonian. 

\begin{figure}[h!]
\begin{center}
\includegraphics[width=0.9\linewidth]{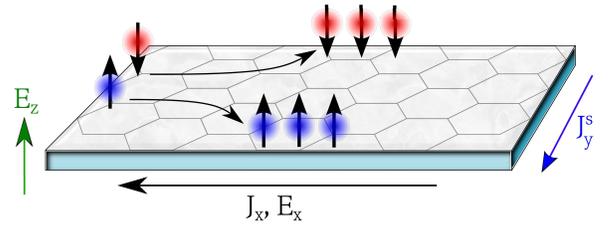}
\caption{\begin{footnotesize}(Color online) Schematic for the spin Hall conductivity.  An electric field $E_x$ is applied to the left such that a charge current $J_x$ is induced in the same direction.  As a result, electrons of different spin flow perpendicularly to the charge current creating a spin imbalance between transverse edges.  This imbalance leads to a finite spin current $J_y^s$.
\end{footnotesize}}\label{fig:HallSchem}
\end{center}
\end{figure}

The spin Hall effect is a phenomenon in which distinct spin species flow to opposite edges of a material resulting in a spin current perpendicular to an applied longitudinal electric field\cite{Hirsch:1999} setting up a spin imbalance as illustrated in Fig.~\ref{fig:HallSchem}.  It has recently been the subject of great interest\cite{Kane:2005, Kane:2005a, Bernevig:2006, Konig:2007, Murakami:2006, Liu:2008, Liu:2008, Duckheim:2009, Abanin:2011, Dyrdal:2012, Tahir:2012, Wenk:2012, Sinova:2004, Pesin:2012} due to its potential application in spintronic technology.  Spintronics hinges on the ability to manipulate an electron's spin to exploit the spin degree of freedom and it relies heavily upon a strong spin-orbit interaction\cite{Kane:2005, Pesin:2012}.  While the long spin relaxation time and length in graphene make it a promising candidate for spintronics\cite{Wang:2012}, silicon has a longer spin-diffusion time\cite{Wang:2012, Appelbaum:2007, Huang:2007} and spin coherence length\cite{Wang:2012, Appelbaum:2007, Huang:2007, Huang:2008, Sanvito:2011} making silicene appear more suitable for spintronic applications\cite{Wang:2012}.
There is an analogous valley Hall effect\cite{Xiao:2007, Rycerz:2007} in which electrons from different valleys (the region about the inequivalent $K$ and $K^\prime=-K$ points of the hexagonal Brillouin zone) flow to opposite transverse edges when the system is subjected to a longitudinal electric field in the presence of intrinsic spin-orbit coupling (SOC).  This effect creates the potential for valleytronic devices.    In silicene, both the spin and valley Hall effect are predicted to be present in the absence of a magnetic field due to the response of the system to an external perpendicular electric field ($E_z$).  

In this paper, we examine the intrinsic spin Hall effect (\emph{i.e.} the effect that arises even in the absence of impurities\cite{Kavokin:2005,Dyrdal:2012,Wenk:2012}) for silicene at finite frequency.  While the DC spin and valley Hall effects in silicene have been examined theoretically\cite{Tahir:2012, Dyrdal:2012} with limited work for the finite chemical potential spin Hall effect\cite{Dyrdal:2012}, we examine the finite frequency response as a function of chemical potential and external electric field, and we provide analytic results for both the spin and valley Hall effects.  We show how the direction of the transverse spin and valley current can be controlled by the frequency of the incident light and by the size of the band gap.  Our paper is organized as follows: in Sec.~II a brief outline of the theory behind silicene is given and analytic formulas are provided.  In Sec.~III, the results for the AC and DC spin and valley Hall conductivities at zero chemical potential are shown with the results for finite chemical potential presented in Sec.~IV.  Our conclusions can be found in Sec.~V.

\section{Theory}

It has been shown\cite{Liu:2011a, Drummond:2012, Ezawa:2012, Ezawa:2012a,Ezawa:2012b} that the low-energy physics of silicene can be captured by the simple tight-binding Hamiltonian
\begin{equation}
\hat H_{K{_\xi}}=\hbar v(\xi k_x\hat\tau_x+k_y\hat\tau_y)-\xi\frac{1}{2}
\Delta_{\rm SO}\hat\sigma_z\hat\tau_z+\frac{1}{2}\Delta_z\hat\tau_z,
\label{Ham}
\end{equation}
where $\tau_i$ and $\sigma_i$ are the Pauli matrices associated with the pseudospin and real spin of the system, respectively. $\xi$ indexes over the two valleys $K$ and $K^\prime$ and can take the values of $\pm 1$, respectively and $v\approx 5\times 10^5$m/s is the Fermi velocity. The first term is the familiar graphene-like Hamiltonian, the second, sometimes referred to as the Kane-Mele term, is related to the strength of the intrinsic SOC, and the final term is associated with the on-site potential ($\Delta_z=E_z d$) resulting from the $A$-$B$ sublattice asymmetry, where $d\approx0.46$\AA\cite{Ezawa:2012b} is the perpendicular distance between the two sublattice planes and $E_z$ is the electric field applied perpendicular to the lattice. In some recent experiments, silicene grown on Ag has been found to have larger buckling values of up to an angstrom.\cite{Kara:2010, Enriquez:2012} For larger buckling, $d$ increases and likewise $\Delta_z$. However, the value of $\Delta_z$ can be changed by adjusting $E_z$. On the otherhand, larger buckling has been shown to increase the spin orbit gap\cite{Liu:2011a}. As $\Delta_{\rm so}$ is an important energy scale in our results, the features shown here will scale accordingly. In particular, depending on the size of $\Delta_{\rm so}$, the phenomena discussed below in this paper will be generated with photon frequencies ranging from THz up to the far- to mid-infrared regime.
   
\begin{figure}[h!]
\begin{center}
\includegraphics[width=0.8\linewidth]{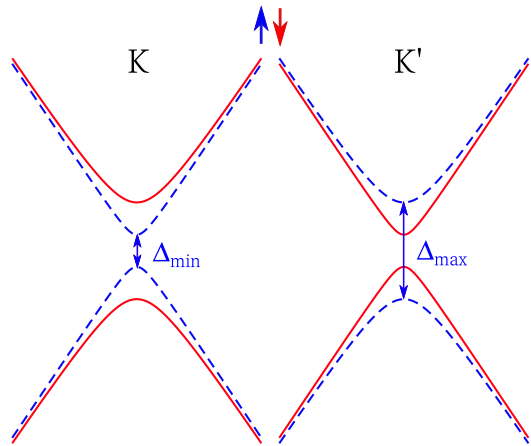}
\caption{\begin{footnotesize}(Color online) Schematic plot of the low-energy band structure of silicene in the presence of a finite electric field.  The spin up bands are given by the dashed blue curves while the spin down bands are represented by the solid red curves.  The blue arrows represent the important energy scales of the system.  The bands are shown for the two valleys $K$ and $K^\prime$.
\end{footnotesize}}\label{fig:Energy}
\end{center}
\end{figure}

In this paper, the first two terms of Eqn.~\eqref{Ham} will be referred to as the Kane-Mele Hamiltonian\cite{Kane:2005a}. In Ref\cite{Ezawa:2012}, a Rashba SOC is also included; however, it is typically neglected\cite{Ezawa:2012} as it is of order 10 times smaller in magnitude than the intrinsic SOC resulting in a negligible effect.  Solving the above Hamiltonian gives the energy dispersion
\begin{equation}
\varepsilon_{\sigma\xi}=\pm\sqrt{\hbar^2v^2k^2+\frac{1}{4}(\Delta_z-\sigma\xi\Delta_{\rm so})^2},
\label{Energy}
\end{equation}
where $\sigma=\pm$1 for up and down spin, respectively.  A schematic plot of the band structure for finite $\Delta_z$ can be seen in Fig.~\ref{fig:Energy} where spin up bands are given by the dashed blue curves and spin down bands by the solid red curves.  For no electric field ($\Delta_z=0$), $\Delta_\text{min}=\Delta_\text{max}=\Delta_{\rm so}$ and there is only one gap between spin degenerate bands which is the result for the Kane-Mele Hamiltonian.

The spin and valley Hall conductivities can be derived via the many body Green's function approach and Kubo formula from Eqn.~\eqref{Ham} as shown in Refs.~\cite{Nicol:2008, Tabert:2012, Stille:2012, Tabert:2013, Tabert:2013a} for electrical charge conductivity and in Ref.~\cite{ZLi:2012} for the spin and valley Hall effects in MoS$_2$.  The Kubo formula for the real part of the conductivity at zero temperature is given by
\begin{align}
\text{Re}\sigma_{\alpha\beta}(\Omega)&=\frac{1}{2\Omega}\int_{|\mu|-\Omega}^{|\mu|}\frac{d\omega}{2\pi}
\int\frac{d^2 k}{(2\pi)^2}\,\\
&\times\text{Tr}\left[\hat{j}_\alpha\hat{A}({\bm k},\omega+\Omega)\hat{j}_\beta\hat{A}({\bm k},\omega)\right],
\label{Kubo}
\end{align} 
where $\mu$ is the chemical potential, $\hat{j}_i$ is the current operator with $i=x$ or $y$, and $\hat{A}(\bm{k},\omega)$ is the spectral function of the Green's function given by $\hat{G}^{-1}(z)=z\hat{I}-\hat{H}$ through the relationship
\begin{equation}
\hat{G}_{ij}(z)=\int_{-\infty}^{\infty}\frac{d\omega}{2\pi}\frac{\hat{A}_{ij}(\omega)}{z-\omega}.
\end{equation}
For spin and valley Hall conductivities, the transverse $\sigma_{xy}(\Omega)$ is required.  The $\hat{j}_x$ operator is given by the usual $\hat{j}_i=e \hat{v}_i$ for electrical conductivity, where $\hbar\hat v_i=\partial \hat H/\partial k_i$.  As in Ref.~\cite{ZLi:2012}, for the spin and valley Hall conductivities, the $\hat{j}_y$ current operator is replaced with $\hat{j}_y^s=\hbar \sigma \hat{v}_y/2$ and $\hat{j}_y^v=\xi \hat{v}_y/2$, respectively.  Evaluation of Eqn.~\eqref{Kubo} and applying Kramers-Kronig relations to obtain Im$\sigma_{xy}(\Omega)$ gives the expressions for the real and imaginary parts of the spin Hall conductivity
\begin{align}\label{SpinHall}
\text{Re}\sigma^{s}_{xy}(\Omega)
&=\sum_{\xi ,\sigma}\xi\sigma\frac{e}{8}\frac{\Delta_{\sigma\xi}}{2\pi\Omega}f(\Omega),\\
\text{Im}\sigma^{s}_{xy}(\Omega)
&=-\sum_{\xi ,\sigma}\xi\sigma\frac{e}{8}\frac{\Delta_{\sigma\xi}}{2\Omega}\Theta(\Omega-\Omega_c),
\end{align}
where $\Delta_{\xi\sigma}=\Delta_z-\sigma\xi\Delta_{\rm so}$, $\Omega_c={\rm max}(2|\mu|,|\Delta_{\sigma\xi}|)$ and
$f(\Omega)=\ln[|\Omega+\Omega_c|/|\Omega-\Omega_c|]$.  $|\Delta_{++}|$ and $|\Delta_{--}|$ are denoted as $\Delta_{\rm min}$ for the smallest band gap and $|\Delta_{+-}|$ and $|\Delta_{-+}|$ as $\Delta_{\rm max}$ for the largest band gap (see Fig.~\ref{fig:Energy}).  The real and imaginary parts of the valley Hall conductivity are
\begin{align}\label{ValleyHall}
\text{Re}\sigma^{v}_{xy}(\Omega)
&=\sum_{\xi ,\sigma}\frac{e}{4\hbar}\frac{\Delta_{\sigma\xi}}{2\pi\Omega}f(\Omega),\\
\text{Im}\sigma^{v}_{xy}(\Omega)
&=-\sum_{\xi ,\sigma}\frac{e}{4\hbar}\frac{\Delta_{\sigma\xi}}{2\Omega}\Theta(\Omega-\Omega_c).
\end{align}

\section{Results for Spin and Valley Hall Conductivity for $\mu=0$}

\begin{figure}[h!]
\begin{center}
\includegraphics[width=1.0\linewidth]{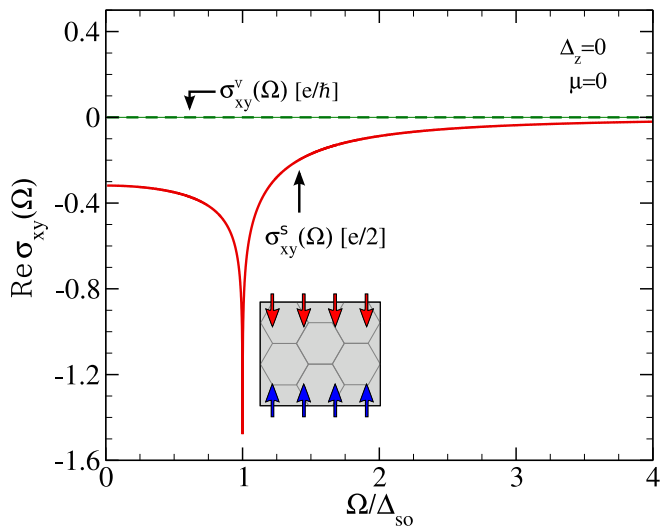}
\caption{\begin{footnotesize}(Color online) Plots of the finite frequency spin Hall (solid red curve) and valley Hall (dashed green curve) conductivities for silicene in the absence of an external electric field ($\Delta_z=0$) and at charge neutrality ($\mu=0$). There is no valley Hall effect; but, the spin Hall effect is sensitive to incident light with a sharp increase in conductivity at $\Omega=\Delta_{\rm so}$.  Inset: to emphasize the character of the spin Hall response, we display a schematic of the spin build up on transverse edges of the sample.
\end{footnotesize}}\label{fig:Hall}
\end{center}
\end{figure}

Plots of the real part of the spin and valley Hall conductivities for charge neutral ($\mu=0$) silicene in the absence of an external electric field ($\Delta_z=0$) are shown in Fig.~\ref{fig:Hall}.
In the DC limit ($\Omega=0$), the expected values for the spin Hall (Re$\sigma^s_{xy}=e/(2\pi)$) and valley Hall (Re$\sigma^v_{xy}=0$) conductivities are obtained\cite{Dyrdal:2012, Tahir:2012}.  If the system is illuminated with photons of frequency $\Omega$, the valley Hall effect remains zero while an increase in the magnitude of the spin Hall response is observed.  When subjected to photons of frequency $\Omega=\Delta_{\rm so}$, a strong response is observed in the spin Hall conductivity.  Thus, a stronger spin Hall response may be accessible in silicene at finite frequency tuned to the spin orbit gap.  For $\Omega$ larger than $\Delta_{\rm so}$, the spin Hall conductivity is quickly diminished and subsequently destroyed at sufficiently high frequency.  As $\Delta_z=0$, these finite frequency results represent those of the Kane-Mele Hamiltonian for a quantum spin Hall insulator and 2D topological insulator.

\begin{figure}[h!]
\begin{center}
\includegraphics[width=1.0\linewidth]{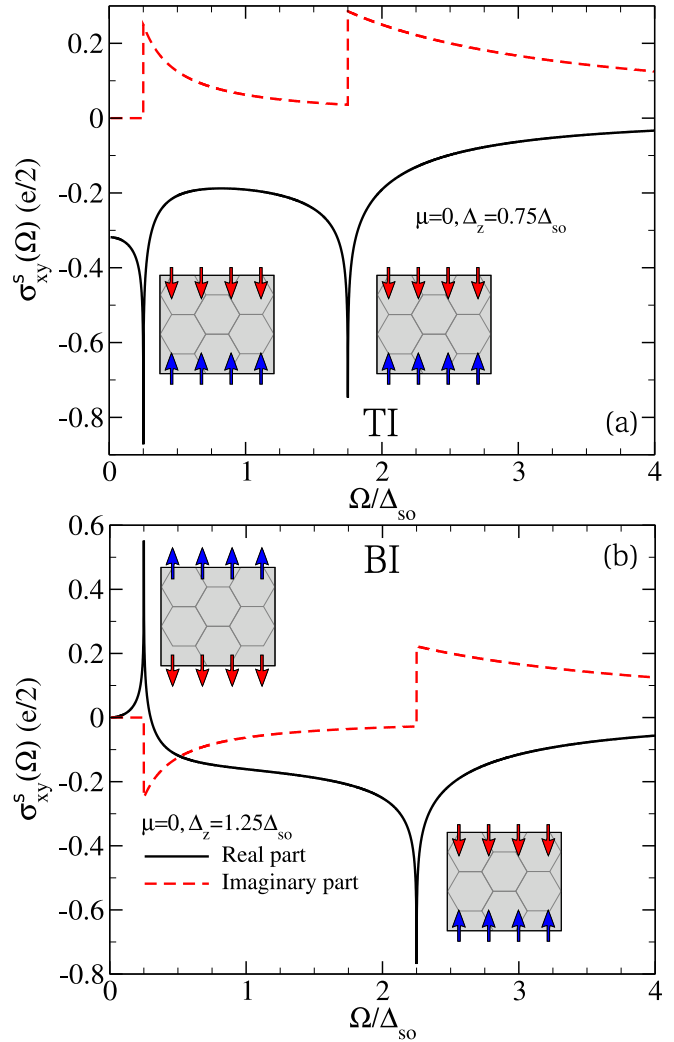}
\caption{\begin{footnotesize}(Color online) The real (solid black) and imaginary (dashed red) parts of the spin Hall conductivity as a function of frequency for $\mu=0$ in (a) the TI regime ($\Delta_z<\Delta_{\rm so}$) for an electric field of strength $\Delta_z=0.75\Delta_{\rm so}$ and (b) the BI regime ($\Delta_z>\Delta_{\rm so}$) for an electric field of strength $\Delta_z=1.25\Delta_{\rm so}$.    In this regime, there is no DC response.
\end{footnotesize}}\label{fig:SpinHall}
\end{center}
\end{figure}

For finite $\Delta_z$, plots of the real (solid black) and imaginary (dotted red) parts of the spin Hall conductivity can be seen in Fig.~\ref{fig:SpinHall}(a) and (b) for the TI ($\Delta_z<\Delta_{\rm so}$) and BI ($\Delta_z>\Delta_{\rm so}$) regimes, respectively.
In reference to Fig.~\ref{fig:SpinHall} and considering the real part, a positive conductivity signifies a net spin up(down) accumulation in one transverse direction while a negative conductivity yields a net spin up(down) accumulation in the opposite direction (schematically shown by the insets).  In the TI regime, Fig.~\ref{fig:SpinHall}(a), the real part of the conductivity (black curve) is always negative, \emph{i.e.} spins of the same orientation always accumulate in the same transverse direction.  The two sharp features in the conductivity are given by the two values of $\Omega_c$ (which are denoted $\Omega_c^{\rm min}$ and $\Omega_c^{\rm max}$ for the first and second feature, respectively).  For $\mu=0$, these represent the onset of the interband transitions at $\Delta_{\rm min}$ and $\Delta_{\rm max}$, respectively.  In the BI regime, Fig.~\ref{fig:SpinHall}(b), there is a sign change in the conductivity which results in a change in spin accumulation as illustrated by the insets.  For low frequency ($\Omega<\Omega_c^{\rm min}$), the spin current flows in the opposite direction to the current in the TI regime while for higher frequencies, the current returns to the direction of the TI regime.  In the TI regime, there is a finite DC response ($\Omega=0$) for Re$\sigma_{xy}^s(\Omega)$ which, for the case of $\mu=0$, is the same value of $e/(2\pi)$ discussed in Fig.~\ref{fig:Hall}. By contrast, in the BI regime there is a zero Hall conductivity at $\Omega=0$.  

\begin{figure}[h!]
\begin{center}
\includegraphics[width=1.0\linewidth]{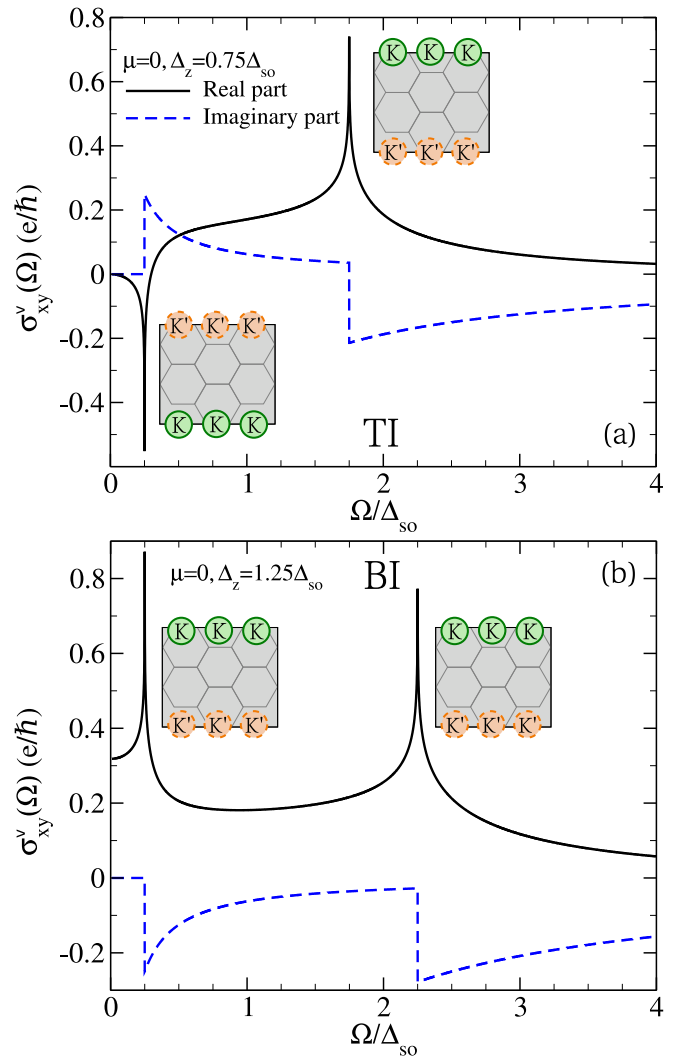}
\caption{\begin{footnotesize}(Color online) The real (solid black) and imaginary (dashed blue) parts of the finite frequency valley Hall conductivity for $\mu=0$ in (a) the TI regime ($\Delta_z<\Delta_{\rm so}$) for an electric field of strength $\Delta_z=0.75\Delta_{\rm so}$.  In this regime, a finite DC response is not present.  (b) The results in the BI regime ($\Delta_z>\Delta_{\rm so}$) for an electric field of strength $\Delta_z=1.25\Delta_{\rm so}$.  Here there is a finite DC response.
\end{footnotesize}}\label{fig:ValleyHall}
\end{center}
\end{figure}

The real (solid black) and imaginary (dashed blue) parts of the valley Hall conductivity can be seen in Fig.~\ref{fig:ValleyHall}(a) and (b) for the TI ($\Delta_z=0.75\Delta_{\rm so}$) and BI ($\Delta_z=1.25\Delta_{\rm so}$) regimes, respectively.
The valley Hall conductivity corresponds to electrons from one valley ($K$ or $K^\prime$) flowing in one direction perpendicular to the longitudinal current while electrons from the other valley flow in the opposite direction creating a net transverse valley imbalance.  In the TI regime, Fig.~\ref{fig:ValleyHall}(a), the conductivity switches sign corresponding to electrons from one valley switching the side to which they flow.  Conversely, in the BI regime, electrons from a given valley flow in only one direction for all $\Omega$.  Again, the sharp features occur at $\Omega=\Omega_c^{\rm min}$ and $\Omega_c^{\rm max}$.  

Examining the combined results for the TI regime shown in Fig.~\ref{fig:SpinHall}(a) and Fig.~\ref{fig:ValleyHall}(a) it is apparent that at $\Omega=\Omega_c^{\rm min}$ electrons of a particular spin and valley label flow to one transverse edge.  At $\Omega=\Omega_c^{\rm max}$, electrons of the same spin but different valley index flow to that transverse edge.  Analogously, in the BI regime (parts (b) of Fig.~\ref{fig:SpinHall} and Fig.~\ref{fig:ValleyHall}) electrons of a particular valley label always flow to one transverse edge while the spin of the electrons changes between $\Omega=\Omega_c^{\rm min}$ and $\Omega_c^{\rm max}$.  This allows one to generate a spin and valley polarized accumulation of carriers on a given transverse edge, \emph{i.e} a build up of a particular spin with either valley label or of a particular valley label and either spin index, depending on the tuning of the incident photon frequency.

\begin{figure}[h!]
\begin{center}
\includegraphics[width=1.0\linewidth]{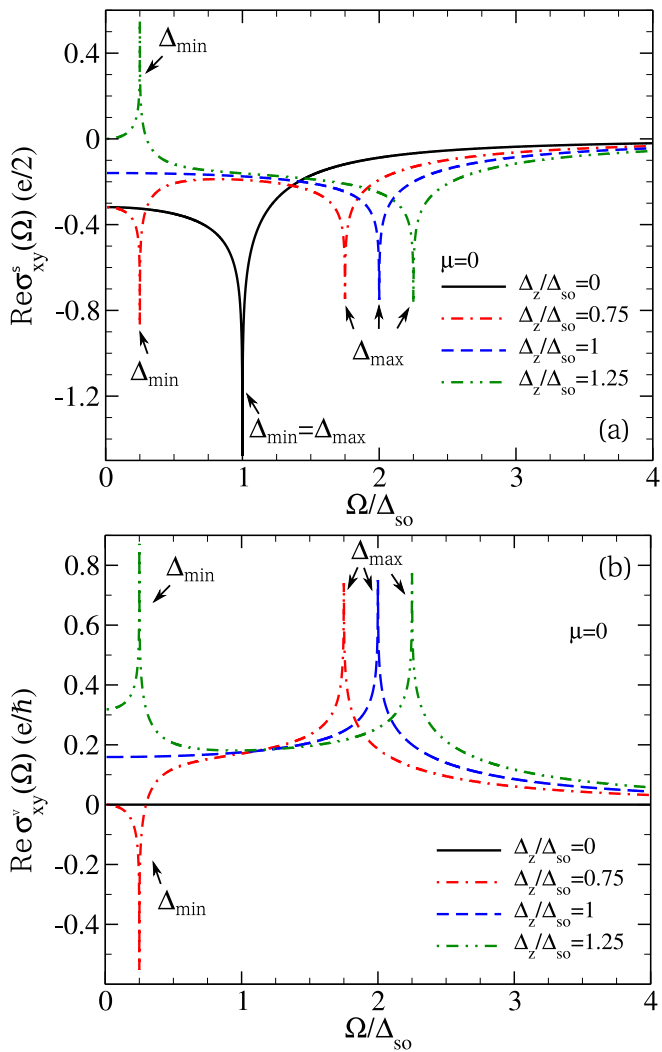}
\caption{\begin{footnotesize}(Color online) The real part of the (a) spin Hall and (b) valley Hall conductivities as a function of frequency for varying electric field strength and $\mu=0$.  In the absence of an electric field (solid black curve) there is only one feature which onsets at $\Omega=\Omega_c^{\rm max}$ for the spin Hall effect and no valley Hall effect. Two features appear in the TI regime (dash dotted red curve) while a single feature of half magnitude appears in the VSPM state (dashed blue curve).  As the system becomes a BI the second feature reappears, however, the lower energy peak switches sign. 
\end{footnotesize}}\label{fig:Hall-Delta}
\end{center}
\end{figure}

The effect of varying the strength of the electric field for fixed $\mu=0$ can be seen in Fig.~\ref{fig:Hall-Delta}(a) and (b) for the spin Hall and valley Hall conductivities, respectively.  As the electric field is increased in the TI regime (dash dotted red curve of Fig.~\ref{fig:Hall-Delta}(a)), the single peak seen in the $\Delta_z=0$ case splits into two, with one peak moving to higher energy and the other to lower energy with increased $\Delta_z$.  In what has been dubbed the valley-spin polarized metal (VSPM) state\cite{Ezawa:2012} ($\Delta_z=\Delta_{\rm so}$) (dashed blue curve) a single feature reappears but of half the strength of the $\Delta_z=0$ case.  This is associated with the lowest gap in the band structure closing leaving only one gapped band at each valley as opposed to the two spin degenerate gapped bands per valley when $\Delta_z=0$.  As the system transitions into the BI regime (dash double-dotted green curve), two peaks reappear as a result of the second gap reopening; however, the low energy interband feature has switched sign as a result of the band inversion that occurs upon moving from the TI to BI regime\cite{Ezawa:2012a}. Both peaks now move to higher energy with a constant separation of 2$\Delta_{\rm so}$ as $\Delta_z$ is increased.  Thus, in the TI regime, the spin Hall effect will lead to spin up(down) accumulation in one transverse direction, while in the BI regime, spin up(down) accumulation can be produced in either direction by tuning the incident photon frequency.  Fig.~\ref{fig:Hall-Delta}(b) shows the corresponding results for the valley Hall conductivity.  The application of an external electric field introduces a finite response with two interband features onsetting at $\Omega_c^{\rm min}$ and $\Omega_c^{\rm max}$.   Here, the switch in direction of the Hall induced imbalance now occurs in the TI regime while it remains the same in the BI regime.  The strong responses in the spin and valley Hall conductivities occur at the same frequencies, $\Omega_c^{\rm min}$ and $\Omega_c^{\rm max}$, and thus a strong spin Hall and valley Hall effect should occur simultaneously. The switch in sign of both effects should allow for a transverse accumulation of charge carriers of particular spin and valley label which can be brought about by tuning the incident frequency.

For $\mu=0$, the DC response for both the spin Hall and valley Hall conductivities have been worked out analytically and are\cite{Dyrdal:2012,Tahir:2012}
\begin{align}
\text{Re}\sigma_{xy}^s(\Omega=0)=\left\lbrace\begin{array}{cc}
\displaystyle -\frac{e}{2\pi}, & \Delta_z<\Delta_{\rm so}\\ \\
\displaystyle -\frac{e}{4\pi}, & \Delta_z=\Delta_{\rm so}\\ \\
\displaystyle 0, & \Delta_z>\Delta_{\rm so}
\end{array}\right.
\end{align}\label{SpinDCm0}
and
\begin{align}
\text{Re}\sigma_{xy}^v(\Omega=0)=\left\lbrace\begin{array}{cc}
\displaystyle 0, & \Delta_z<\Delta_{\rm so}\\ \\
\displaystyle\frac{e}{4\pi\hbar}, & \Delta_z=\Delta_{\rm so}\\ \\
\displaystyle\frac{e}{2\pi\hbar}, & \Delta_z>\Delta_{\rm so}
\end{array}\right. .
\end{align}\label{ValleyDCm0}
Thus, a simultaneous DC response for both the spin and valley Hall conductivities is only attained in the VSPM state.  Unlike the $\Delta_z=0$ case where a finite DC spin Hall effect exists but no DC valley Hall effect is present, a finite DC valley Hall conductivity is obtained in the BI regime; however, the DC spin Hall conductivity is now zero.  Our results at finite frequency properly  capture this limiting behaviour as $\Omega\rightarrow 0$.

\section{Results for finite chemical potential $\mu$}

\begin{figure}[h!]
\begin{center}
\includegraphics[width=1.0\linewidth]{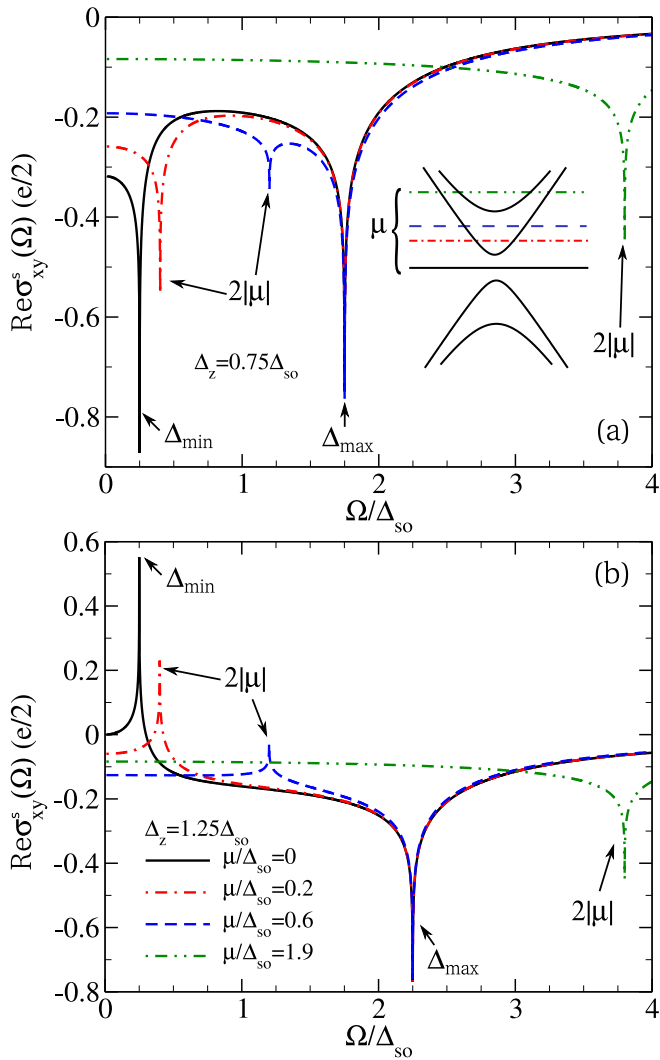}
\caption{\begin{footnotesize}(Color online) Real part of the spin Hall effect for finite chemical potential in the (a) TI regime for $\Delta_z=0.75\Delta_{\rm so}$ and (b) BI regime for $\Delta_z=1.25\Delta_{\rm so}$.  As shown in (a), the spin imbalance remains in the same direction in the TI regime; however, as $\mu$ is increased such that $2|\mu|>\Delta_{\rm min}$, the first feature sits at $\Omega=2|\mu|$ and decreases in strength while the second peak remains unchanged.  As $2|\mu|$ becomes greater than $\Delta_{\rm max}$ only one feature is present and it onsets at $2|\mu|$.  The DC response remains consistent with Eqn.~\eqref{DCSpinHall}.  In (b), the same behaviour is seen as in (a) in regards to the onset and strength of the features.
\end{footnotesize}}\label{fig:SpinHall-mu}
\end{center}
\end{figure}

The spin Hall effect for varying chemical potential can be seen in Fig.~\ref{fig:SpinHall-mu}(a) and (b) for the TI and BI regime, respectively.  
The finite-$\mu$ results for the valley Hall effect are not shown as they are analogous.  For $\mu$ in the energy gap (\emph{i.e.} $2|\mu |<\Delta_{\rm min}$), the system behaves as in the undoped case.  When $\mu$ is increased such that it sits in the lowest conduction band ($\Delta_{\rm min}<2|\mu |<\Delta_{\rm max}$) (dashed dotted red and dashed blue curves) the location of the first peak occurs at $2|\mu|$ as the first interband transitions can only onset at this value due to Pauli blocking.  Finally, if the chemical potential is placed through both conduction bands ($2|\mu |>\Delta_{\rm max}$) (dashed double-dotted green curve), only one feature at $\Omega=2|\mu|$ is present.  Note that as $\mu$ is increased, the strength of the overall response diminishes with exception of that associated with $\Delta_{\rm min}$ and $\Delta_{\rm max}$ should either peak not be Pauli blocked.

In general, accounting for a finite chemical potential, the DC response for the real part of the spin Hall conductivity can be worked out analytically and is
\begin{align}\label{DCSpinHall}
\text{Re}&\sigma_{xy}^s(\Omega=0)\notag\\
&=\left\lbrace\begin{array}{cc}
\displaystyle\frac{e}{4\pi}\bigg[\text{sgn}(\Delta_{++})-\text{sgn}(\Delta_{+-})\bigg], & \Delta_{\rm min}>2|\mu|\\
\displaystyle\frac{e}{4\pi}\bigg[\frac{\Delta_{++}}{2|\mu|}-1\bigg], & \Delta_{\rm min}<2|\mu|<\Delta_{\rm max}\\
\displaystyle-\frac{e}{4\pi}\frac{\Delta_{\rm so}}{|\mu|}, & \Delta_{\rm max}<2|\mu|,
\end{array}\right.
\end{align}
where sgn$(x)$=1 for $x>0$, 0 for $x=0$ and -1 for $x<0$ which is in agreement with the results shown in Refs.~\cite{Dyrdal:2012, Tahir:2012} for finite $\mu$.  The DC response for the real part of the valley Hall conductivity can also be worked out analytically and is given by
\begin{align}\label{DCValleyHall}
\text{Re}&\sigma_{xy}^v(\Omega=0)\notag\\
&=\left\lbrace\begin{array}{cc}
\displaystyle\frac{e}{4\pi\hbar}\bigg[\text{sgn}(\Delta_{++})+\text{sgn}(\Delta_{+-})\bigg], & \Delta_{\rm min}>2|\mu|\\
\displaystyle\frac{e}{4\pi\hbar}\bigg[\frac{\Delta_{++}}{2|\mu|}+1\bigg], & \Delta_{\rm min}<2|\mu|<\Delta_{\rm max}\\
\displaystyle\frac{e}{4\pi\hbar}\frac{\Delta_{z}}{|\mu|}, & \Delta_{\rm max}<2|\mu|,
\end{array}\right. .
\end{align}
Therefore, unlike the undoped case, with a finite chemical potential $|\mu|>\Delta_{\rm min}/2$, a non-zero DC response for both spin and valley Hall conductivities is possible in all insulating regimes.  Note that for $2|\mu|>\Delta_{\rm max}$, the DC spin Hall effect is controlled by $\Delta_{\rm so}$ whereas the DC valley Hall response is controlled by $\Delta_z$.  Thus, for large chemical potential, the spin Hall effect is an intrinsic property of the system while the valley Hall effect is generated and tuned by an external electric field.

Note that electron-electron and electron-phonon interactions have been neglected in our work. For the dynamical conductivity of 2D crystals some effects of the electron-phonon interaction are illustrated in Refs.\cite{ZLi:2012} and \cite{Carbotte:2010}. Electron-electron interactions have been suggested to give a frequency dependent scattering rate in graphene, along with features due to plasmaron formation\cite{Carbotte:2012, Leblanc:2011}. The substrate is particularly important in adjusting the strength of the electron-electron interactions. Our calculations assume that such effects are small or can be made to be so by suitable choice of substrate or making the material freestanding. Optical experiments in graphene have confirmed that the single-particle picture works well for capturing the essence of the finite frequency response.

\section{Conclusions}

The finite-frequency spin and valley Hall conductivities of silicene, germanene and similar 2D crystals have been calculated.  Analytic results are presented for the finite frequency and DC spin and valley Hall conductivities which are in agreement with what has previously appeared in the literature.  The type of insulating phase of the system is shown to play an important role in the direction of the spin and valley accumulation.  Indeed, by tuning the frequency of incident photons, a transverse spin and valley polarized accumulation of carriers can be obtained.  Thus, a build up of either spin and valley label can be generated on a transverse edge of the system.  The frequency for onset of the strong spin and valley Hall responses can be controlled by varying the strength of the electric field and chemical potential.  While for charge neutrality there is a finite DC spin Hall effect in the absence of an electric field, the application of a perpendicular electric field is required to allow for a finite DC valley Hall conductivity.  In the VSPM phase, both a finite spin Hall and valley Hall response is obtained in the DC limit.  The DC values can also be tuned by the perpendicular electric field strength and chemical potential.  As samples of silicene become more readily available we trust that silicene will show great promise in the development of spin- and valleytronic devices.

\begin{acknowledgements}
We thank D. A. Abanin for helpful discussions.  This research was supported by the Natural Science and Engineering Research Council of Canada.
\end{acknowledgements}

\bibliographystyle{apsrev4-1}
\bibliography{svhall}

\end{document}